\documentclass{aa}
\usepackage{graphicx}
\usepackage{txfonts}
\usepackage{lipsum}

\newcommand{\gc}{$\gamma$\,Cas}
\newcommand{\kms}{km\,s$^{-1}$}

\begin{document}

\title{Orbital motion detected in \gc\ Fe K emission lines}

\author{Ya\"el Naz\'e\inst{1}\fnmsep\thanks{F.R.S.-FNRS Senior Research Associate}
     \and Masahiro Tsujimoto\inst{2,3}
     \and Gregor Rauw\inst{1}
     \and Sean J. Gunderson\inst{4}
        }

   \institute{GAPHE, STAR, Universit\'e de Li\`ege, B5c, All\'ee du 6 Ao\^ut 19c, B-4000 Sart Tilman, Li\`ege, Belgium
              \email{ynaze@uliege.be}
   \and  Institute of Space and Astronautical Science, JAXA, 3-1-1 Yoshino-dai, Chuo-ku, Sagamihara, 252-5210, Kanagawa, Japan
   \and Dep. of Astronomy, Graduate School of Science, The University of Tokyo, Hongo 7-3-1, Bunkyo-ku, 113-0033, Tokyo, Japan
   \and Kavli Institute for Astrophysics and Space Research, M.I.T., 77 Massachusetts Avenue, Cambridge, MA 02139, USA
   }

   \abstract{A subset of Be stars, typified by the naked-eye star \gc, exhibits unusually bright and hard X-ray emission, the origin of which has remained debated for five decades. We performed high-resolution X-ray spectroscopic monitoring of \gc\ with the {\it Resolve} instrument aboard the X-Ray Imaging and Spectroscopy Mission (XRISM). X-ray lines from the ultra-hot plasma and fluorescence from cooler material exhibit Doppler shifts consistent with orbital motion, not of the Be star itself, but of its low-mass companion (previously shown to be a white dwarf). This first evidence of orbital motion for the hard X-ray emitting plasma uniquely links it to the scenario of accretion onto the white dwarf companion. The modest line broadening further indicates that fluorescence occurs on the white dwarf surface and excludes X-ray generation in the inner parts of an accretion disc. Our findings identify \gc\ and its analogues as the previously elusive, but long predicted class of binaries composed of a Be and a white dwarf. Identifying the origin of the hard X-rays from \gc\ and its analogues, which represent about 10\% of early-type Be stars, provides a key input for population synthesis models of massive binary evolution.}

   \maketitle
   
\section{Introduction}
Stars are sources of X-rays, and massive stars -- those born with masses exceeding ten solar masses (10\,M$_{\odot}$) -- are no exception \citep{rau22book}. In such stars, high-energy emission is typically associated with strong shocks embedded in their stellar winds \citep{fel97,krt09}. These dense outflows are driven by intense stellar ultraviolet radiation, a process that is inherently unstable and naturally gives rise to shocks that heat the plasma to moderate temperatures (about 5\,MK). Additional thermal X-ray emissions, often at somewhat higher temperatures, can arise from colliding wind flows in massive binaries \citep{ste92} or from wind confinement by strong magnetic fields \citep{udd14}. 
  
Beyond their powerful winds, another key characteristic of massive stars is high multiplicity \citep{moe17}. The ubiquitous presence of companions has significant implications for stellar evolution. Binary interactions can notably strip one star of its outer layers while transferring mass and angular momentum to its companion. This process is now widely considered as the primary formation channel for Be stars (e.g. \citealt{kle19}), a class of rapidly-rotating B-type stars surrounded by circumstellar discs \citep{riv13}. These discs, which are in Keplerian rotation and are episodically replenished by localised stellar mass ejections \citep{lab25}, produce characteristic optical emission lines that account for the `e' designation in the spectral type. Because of their binary origin, Be stars serve as crucial laboratories for testing models of binary stellar evolution.

Evolutionary models predict that companions of Be stars should include stripped helium stars, white dwarfs, neutron stars, and black holes \citep{sha14}. Observational studies have identified numerous Be X-ray binaries hosting neutron stars \citep{rei11} and have detected about a dozen hot stripped helium star companions through their ultraviolet signatures \citep{wan21}. However, the numerous predicted systems composed of a Be star and a white dwarf remain a missing population.

Indeed, the few previously proposed detections remain debated. For example, earlier studies attributed an excess in the extreme UV range detected in some B stars to a (hot) white dwarf  \citep{bur98,bur99,bur00,ber00,hol13}, but further investigation of at least one of them ($\lambda$\,Sco) revealed that it did not harbour any white dwarf \citep{uyt04}. In parallel, studies have identified some supersoft X-ray sources, usually associated with white dwarfs, as corresponding to Be counterparts, leading to the hypothesis that both objects form a binary \citep{kah06,ori10,li12,stu12,cra18,hab20,coe20,ken21,gau25,mar25}. However, \citet{sch25} express strong doubts about the secure, unequivocal association between supersoft X-rays and white dwarfs, especially in the absence of a (recorded) nova eruption. Finally, observations identified that Be stars appear near the location of two peculiar novae, and the periodic photometric variations were interpreted as ellipsoidal variations of the Be star in its orbit around the flaring white dwarf \citep{cha25,sch25}. However, the optical photometry was not fitted to assess the plausibility of the ellipsoidal interpretation, and no radial velocity curve exists, leaving some doubt about the (physical) association of the two stars. The bulk of the Be-white dwarf systems thus remains to be identified.

 \gc\ was the first object identified as a Be star \citep{sec66}. It is a known binary system with a circular orbit and a period of 203\,d \citep{har00,nem12,smi12}. It was considered a typical Be star until 1975, when \citet{mas76} identified that it exhibited unusual high-energy properties. Subsequent studies revealed that its X-ray spectrum is dominated by thermal emission from an extremely hot ($\sim$150\,MK), optically thin plasma \citep{smi16}. No trace of significant non-thermal emission was found \citep{shr15}. At such high temperatures, lines from highly ionised iron (Fe\,{\sc xxv} at 6.7\,keV and Fe\,{\sc xxvi} at 7\,keV) become prominent. Moreover, these lines appear flanked by a fluorescence iron line at 6.4\,keV, corresponding to X-rays being absorbed and re-emitted by cooler material in close proximity to the X-ray source -- another feature that is unique amongst massive stars. Additionally, \gc\ is approximately 40 times more X-ray luminous than is typically expected for massive stars. Its X-ray flux varies by a factor of a few on timescales as short as a few seconds \citep{lop10,smi16}, again in stark contrast to the steady X-ray output of massive stars on these timescales. Observations also reveal variations on longer timescales, including transient changes in absorption towards the warm plasma \citep{ham16,smi19,rau22} and overall changes in flux, notably correlated with broadband photometric changes but not with orbital phase or H$\alpha$ emission strength \citep{mot15,rau22}. 

Over the past two decades, sensitive X-ray observatories have revealed similarly anomalous X-rays in two dozen early-type Be stars (e.g. Table 5 of \citealt{naz20}), thereby establishing a subclass of sources called `\gc\ analogues' (a term coined by \citealt{smi99}). 

Several hypotheses have been proposed to explain this peculiar high-energy emission. One scenario attributes the X-rays to magnetic reconnection events between small-scale magnetic loops at the Be star's surface and the toroidal field of its surrounding disc \citep{smi98,mot15}. Although current spectropolarimetric surveys rule out the presence of large-scale magnetic fields in Be stars, small-scale fields -- potentially generated in subsurface convection zones \citep{can11} -- remain undetectable and theoretically plausible.

An alternative class of scenarios attributes the X-ray emission not to the Be star itself but to a binary companion. The nature of this companion critically shapes the interpretation. If it is a stripped hot star, the collision of its wind with the Be disc could, in principle, generate X-rays \citep{lan20}. However, this model appears inconsistent with the observed properties of stripped stars, colliding winds, and X-rays from Be-stripped star systems \citep{naz22}. If the companion is a neutron star, the observed X-rays could only be reproduced during a short-lived `propeller' phase of accretion (which occurs for fast-rotating magnetic neutron stars: direct accretion is prohibited, and X-rays are produced in a surrounding shell of hot material; see \citealt{pos17}). However, the rarity of this phase and the discrepancies between the observed and predicted X-ray properties make this explanation unlikely \citep{smi17,rau24}. 

This leaves an accreting white dwarf as the only possible candidate in this category \citep{mur86,app02,ham16,tsu18,toa25,gun25}. The X-ray signatures of \gc\ analogues -- namely high plasma temperatures, the presence of a 6.4\,keV fluorescence feature, and rapid, short-term variability -- closely resemble those observed in known accreting white dwarf systems such as cataclysmic variables and symbiotic stars. However, the analogy is imperfect, since no known accreting white dwarf replicates the full set of \gc\ characteristics \citep{naz24}. This may reflect the vastly different configurations of the known cases \citep{muk17}: extremely short periods ($<$2\,d) in cataclysmic variables (rather than several tens or hundreds of days in \gc\ analogues) and white dwarfs paired with low-mass red giant stars in symbiotics (rather than with massive stars). Material feeding the white dwarf disc comes from the primary star's wind or from the star overfilling its Roche lobe in such systems, rather than from a decretion disc as proposed for \gc\ analogues. Nonetheless, generic spectral models for accreting white dwarfs provide good fits to the broadband X-ray spectra of \gc\ analogues \citep{tsu18,toa25,gun25}, although alternative models yield comparably good fits (e.g. \citealt{smi04,lop10,rau22}).

A final piece of supporting evidence is the lack of detectable optical signatures for the companions in some studied \gc\ analogues, including \gc\ itself \citep{kle24}. This favours a white dwarf interpretation for the companion over a more luminous stripped star. Such a companion is required for an accreting white dwarf scenario, but models in which X-rays are produced near the Be star can accommodate any type of companion, including white dwarfs. To date, observational evidence remains inconclusive as to whether the high-energy emission in \gc\ analogues arises from the Be star itself or from its companion. The answer lies in a detailed analysis of the iron X-ray lines, because highly ionised lines track the primary source of the peculiar X-rays, while fluorescence lines probe that source as well as the cool illuminated material. While the precision of X-ray facilities previously remained too low, such a study is now possible thanks to the X-ray Imaging and Spectroscopy Mission (XRISM) and its high-resolution spectrometer {\it Resolve}. This paper reports the results of our XRISM monitoring campaign dedicated to \gc.

\section{Observations with XRISM}
The XRISM observatory \citep{tas25} orbits the Earth at an altitude of 570\,km, corresponding to a period of $\sim$100\,minutes and an orbital inclination angle of 31$^{\circ}$. Targets are only observed during intervals when XRISM's line of sight is unobstructed by Earth. The {\it Resolve} onboard spectrometer is equipped with a 6$\times$6 microcalorimeter array \citep{kel25,ish25}, which is capable of measuring the energy of individual incoming photons ($\sim$1\,fJ) through extremely small temperature increases ($\sim$0.001\,K) to achieve a relative line energy determination with an accuracy of 30\,ppm \citep{por25}. 

\begin{figure}
  \begin{center}
    \includegraphics[width=8cm]{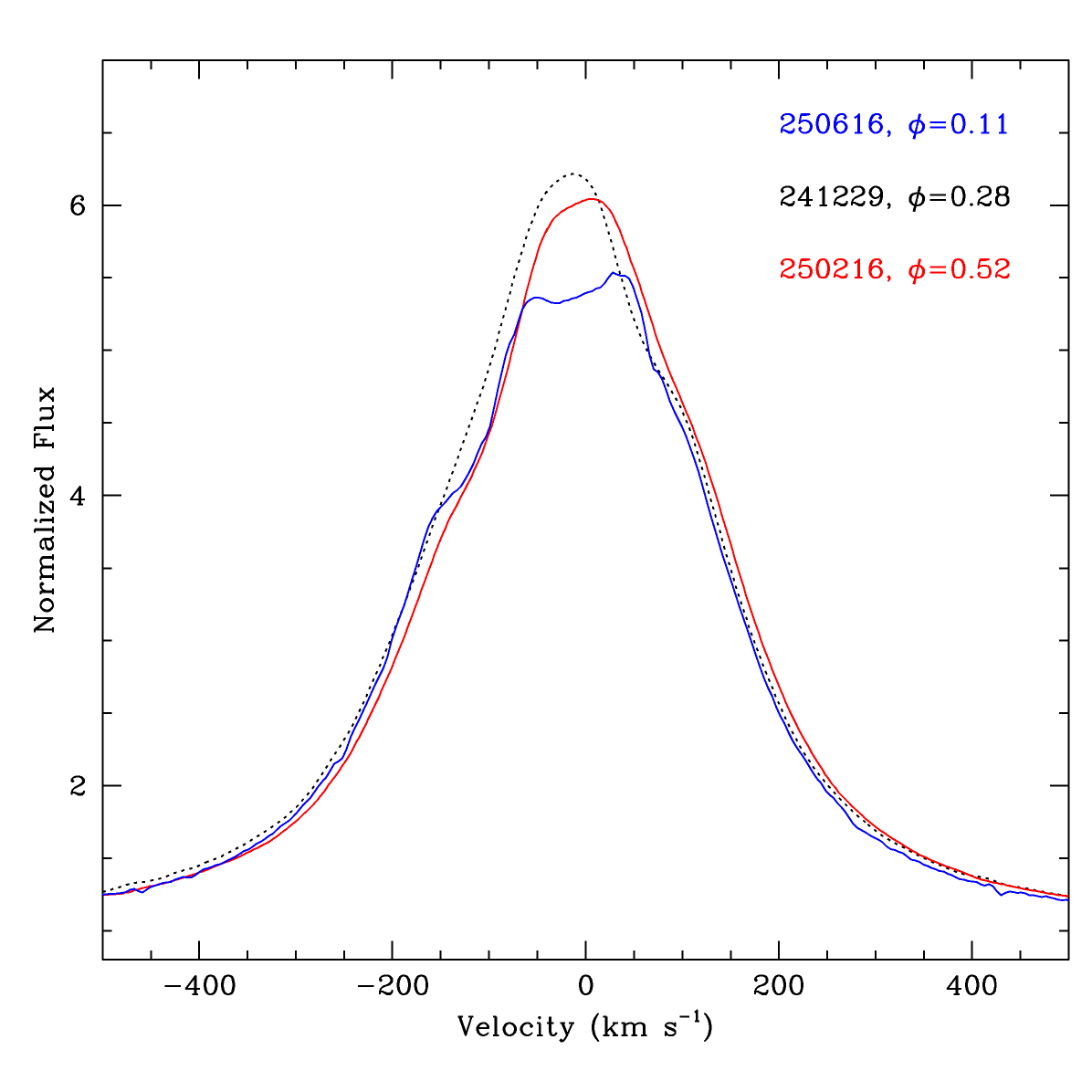}
  \end{center}  
  \caption{H$\alpha$ line during  XRISM observations \citep{naz25}. Because the emission arises from the decretion disc surrounding the Be star, it shifts with the Be star's orbital motion.  \label{fig:optlines}}
\end{figure}

We conducted observations of \gc\ at three distinct orbital phases (Table \ref{tab:rv}): near the quadrature, when the Be star exhibits minimal radial velocity ($\phi$=0.11, June 2025); near conjunction, with the Be star in front of its companion ($\phi$=0.28, December 2024); and near the opposite quadrature ($\phi$=0.52, February 2025). This temporal sampling spans the full range of orbital motion, greatly enhancing the diagnostic power of the data. Simultaneous optical observations confirmed that the Be disc was well developed at all three epochs, enabling a consistent phase-by-phase comparison (\citealt{naz25} and Fig. \ref{fig:optlines}).

The {\it Resolve} data were initially processed using the standard XRISM pipeline \citep{doy22}, followed by an additional filtering step based on the pulse shape of the detected X-ray events \citep{moc25}. We excluded events from pixel 27 in all three observations due to anomalous behaviour. For the June 2025 observation, we also removed events from pixel 7 because of elevated noise levels. Both of these pixels lie at the periphery of the detector array and contributed only a small fraction of the total counts; therefore, their removal has a negligible impact on the signal-to-noise ratio. For the December 2024 observations, a very short telemetry loss occurred for pixel 17, requiring a manual merging of two unintentionally divided time intervals to reconstruct the gain variation curve of that pixel (see below). The resulting net count rate of \gc\ is 2.5~cts~s$^{-1}$ in the 2--12\,keV band. At this rate, the source photons dominate over the very low instrumental background, and a large fraction (95\%) of the events are also individually detected within a single pixel, allowing for highly accurate energy measurements. 

The energy calibration of {\it Resolve} varies with time, primarily due to variations in the cold-stage temperature (following sub-Kelvin cooler recycling) and in the satellite's orientation with respect to the Sun and Earth. These effects result in changes in the energy gain (pulse height response to a monochromatic X-ray energy), inducing energy shifts of up to 5\,eV at 5.9\,keV, equivalent to a velocity uncertainty of $\sim$250\,\kms. To monitor and correct these shifts, {\it Resolve} is equipped with three onboard calibration systems. 

The first consists of modulated X-ray sources \citep{shi25}, which we did not use in the current observations. The second system includes five $^{55}$Fe radioactive sources mounted on the filter wheel atop the cryostat, where the detector is installed. During Earth occultations, the filter wheel occasionally rotates into position, illuminating the full detector array. These exposures occurred two to three times during each observation, as well as before and after science exposures. When this system is used, the energy gain of individual pixels can be tracked using lines of Mn K$\alpha$ (5.90\,keV) produced after electron capture decay. Pixel-by-pixel gain variation over time is then determined from a polynomial fit to data taken during the calibration exposures, and the derived correction is applied to each event as part of the standard XRISM-{\it Resolve} data reduction pipeline. 

The third calibration source is a dedicated $^{55}$Fe source that continuously illuminates pixel 12, which is located outside the main array. This source provides a continuous temporal monitoring of the energy calibration and allows estimation of the potential bias due to differing illumination conditions during calibration and science exposures (i.e. inside versus outside of Earth occultation). Representative values for this global velocity bias are computed for each observation (see Fig. 5 in the {\it Resolve} quality reports\footnote{https://heasarc.gsfc.nasa.gov/docs/xrism/analysis/gainreports/index.html}, version 11). They are listed in Table \ref{tab:rv} and allowed us to account for the differential biases among the observations. We applied an additional correction to account for Earth's motion around the Solar System barycentre. We calculated its values for the line of sight towards \gc\ at every observing date using the IRAF and MIDAS public software packages (see Table \ref{tab:rv}). We added these two independent corrections to the velocity values derived from the spectral fits (see the next section). 

In principle, another correction should be applied to account for the velocity of XRISM in its low-Earth orbit. The exact spacecraft velocity at any time during an observation is available through the spacecraft orbit file provided to the users. However, in our case, the average spacecraft velocity projected along the line of sight is effectively zero when integrated over an entire exposure, with a dispersion $<$5\,\kms. This is negligible compared to other sources of uncertainties, so we did not correct for it. Overall, the calibration strategy provides an energy-scale accuracy of $\sim$0.2\,eV at 5.9\,keV \citep{por25}, corresponding to a velocity precision of 10\,\kms.  

\section{Data modelling}

\begin{figure}
  \begin{center}
    \includegraphics[width=8cm]{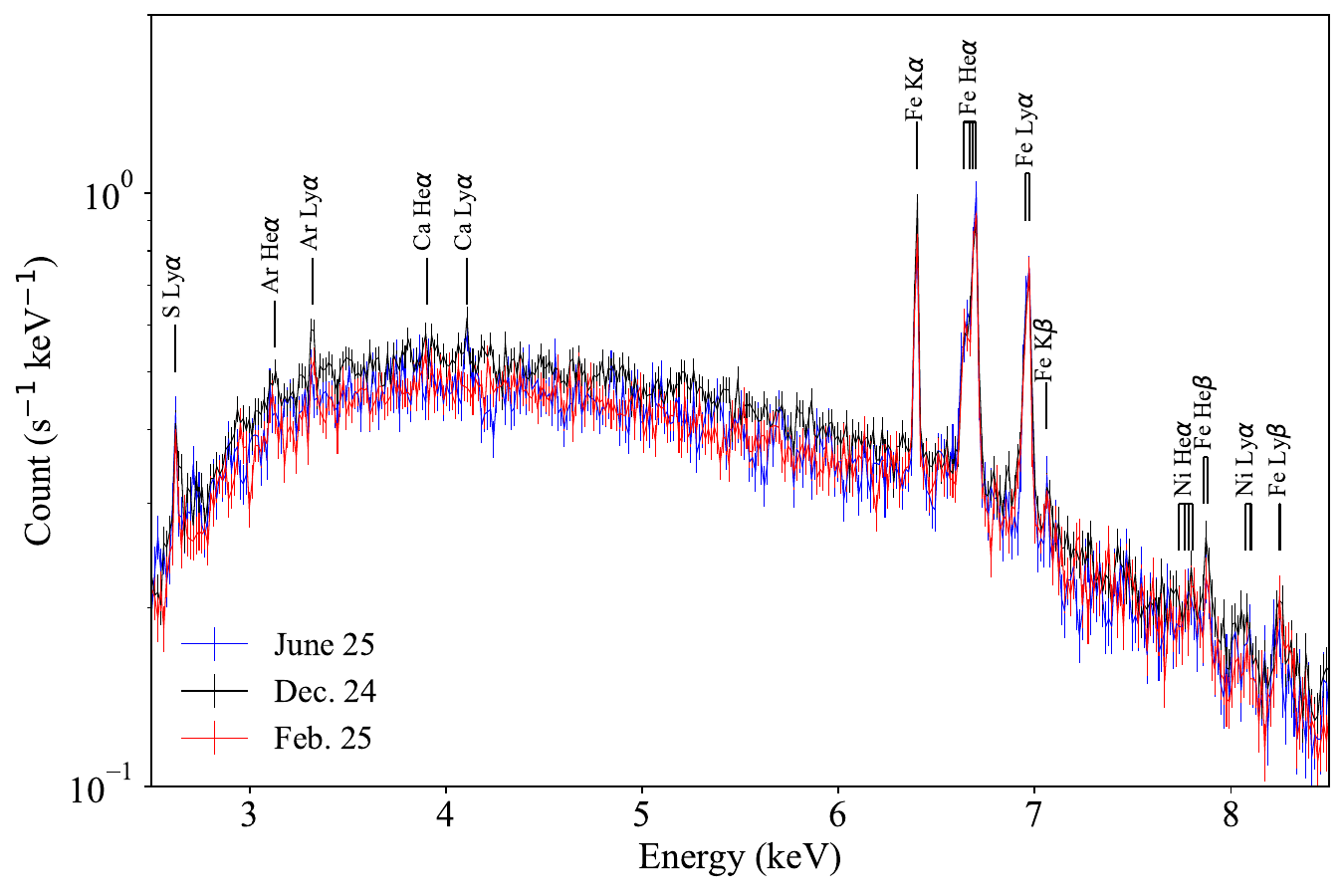}
  \end{center}  
  \caption{{\it Resolve} spectra over the full spectral range (7.5\,eV per bin or 15-channel grouping). The three components of the iron complex at 6.4--7.0\,keV are prominent, and additional fainter lines of other metals are also visible.   \label{fig:allspec}}
\end{figure}

\begin{table*}
  \scriptsize
  \caption{Observation list and line-fitting results.
 \label{tab:rv}}
  \begin{tabular}{lcccccccccccc}
    \hline
Short  &  ObsID & Date (T$_{exp}$ in ks) & $HJD$ & $\phi$ & Cor$_c$ & Cor$_b$ & \multicolumn{2}{c}{Fluorescence} &  \multicolumn{4}{c}{Fe\,{\sc xxv} and Fe\,{\sc xxvi} lines}\\
name  & 20100+ &                 & $-2460000.$ &        & (\kms)  & (\kms) & shift$^f$ & $\sigma^f_G$ & shift$^i$ & $\sigma^i_G$ & $kT^i$ & A$^i_{\rm Fe}$ \\
& & & & & &          &(\kms)     & (\kms)    & (\kms)   & (\kms)    & (keV) & \\
\hline
June 25 & 201 & 2025-06-16T23:46:54 (46.5)& 843.488 & 0.11 &  +0.6 & +12.9 & 175$\pm$21 & 170$\pm$28 & 121$\pm$19 & 366$\pm$23 & 8.22$\pm$0.22 & 0.234$\pm$0.012\\
Dec 24  & 301 & 2024-12-29T14:16:44 (34.4)& 674.097 & 0.28 & +10.3 &--16.2 & 128$\pm$32 & 244$\pm$42 & 123$\pm$23 & 414$\pm$30 & 8.31$\pm$0.24 & 0.269$\pm$0.016\\
Feb 25  & 401 & 2025-02-15T19:45:34 (59.3)& 722.322 & 0.52 &--12.2 &--19.4 &  27$\pm$18 & 177$\pm$20 &  34$\pm$23 & 495$\pm$32 & 8.43$\pm$0.19 & 0.236$\pm$0.012\\
\hline
  \end{tabular}
  \tablefoot{The second column provides the observation identifier; the third, the date at mid-exposure (exposure length in parentheses); and the next two columns, the corresponding heliocentric Julian date and orbital phase $\phi$ computed from the recent ephemerides of \citet{naz25}. The following two columns give the corrections due to refined calibration based on Mn\,K$\alpha$ and to Earth motion around the Sun. The next two columns provide the shift and Gaussian broadening measured on spectra with five-channel binning using the neutral iron model with seven components (see text for details). The last columns yield the shift, Gaussian broadening, temperature, and iron abundance (with respect to hydrogen and to the solar ratio) derived from fitting the highly ionised iron lines with a thermal plasma model (see text for details). Velocity values include the required corrections and all provided errors are $\pm1\sigma$ uncertainties.}
\end{table*}

We combined the high-quality events to construct an X-ray spectrum for each of the three observations. The resulting spectra prominently show the iron complex, with the fine structure in these lines resolved for the first time (Figs. \ref{fig:allspec} and \ref{fig:linesx}). We also detect mission lines from other elements (S, Ar, Ca, and Ni) and higher-order transitions of iron (e.g. Fe\,{\sc xxvi} Ly$\beta$). However, given their comparatively low signal-to-noise ratio, these lines do not provide the precision necessary to achieve our goals. We only note that the K$\beta$ fluorescence line has a strength of  $\sim$12\% of that of K$\alpha$, as expected \citep{pal03}. Consequently, the analysis presented here focuses on the well-exposed iron complex at 6.4-7.0 \, keV. We carried out spectral modelling using the the X-ray spectral fitting package XSPEC, with three independent binnings: (1) unbinned data, (2) fixed binning with five energy channels (or 2.5 \, eV) per bin, and (3) optimal channel grouping calculated with the tool \texttt{ftgrouppha} \citep{kaa16}. 

The iron fluorescence near 6.4 keV follows the photoionisation of a K-shell electron of low-ionisation states of iron (less than Fe\,{\sc x}). The ensuing gap on the K-shell is filled by an L-shell electron which sheds its excess energy by emitting a fluorescent line photon in about one third of all cases. The fluorescence feature near 6.4\,keV consists of a blend of several subcomponents (25 for Fe\,{\sc i} or 16 for Fe\,{\sc viii}; \citealt{pal03}). They are usually grouped into two unresolved transition arrays named K$\alpha_1$ and K$\alpha_2$. For solid-state Fe\,{\sc i} target atoms, the measured lines can also be modelled with seven Lorentzian components, often broadened to account for the grouping of nearby lines \citep{hol97}. We modelled the fluorescence complex of \gc\ using three different approaches: (1) the Xspec model \texttt{gsmooth*zfeklor}, where \texttt{gsmooth} applies Gaussian broadening and {\texttt{zfeklor} incorporates the seven solid-state Fe\,{\sc i} component model of \citet{hol97} with fixed line ratios and a global velocity shift as a free parameter, (2) a two-component Gaussian model representing the main K$\alpha_1$ and K$\alpha_2$ components of Fe\,{\sc i} or Fe\,{\sc viii} \citep{pal03}, assuming identical line widths and velocity shifts but leaving the line ratio free to vary, (3) a model using the full set of subcomponents predicted  by \citet{pal03} for Fe\,{\sc i} or  Fe\,{\sc viii}, assuming identical line widths and velocity shifts but fixed line ratios. In all cases, we modelled the continuum using a power law with a zero slope. 

\begin{figure}
  \begin{center}
    \includegraphics[width=8cm]{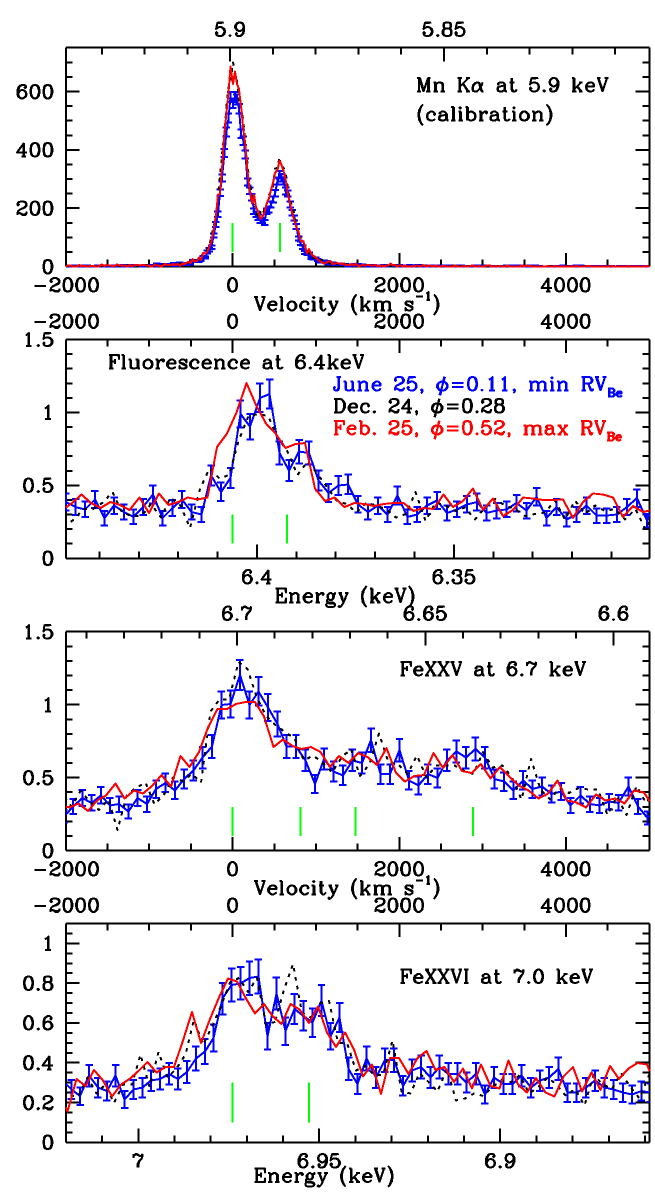}
  \end{center}  
  \caption{Observed profile of the three iron features in the \gc\ spectrum: fluorescence at 6.4\,keV (second panel) and Fe\,{\sc xxv} and Fe\,{\sc xxvi} contributions (third and fourth panels, respectively). For comparison, the stationary calibration line of Mn K$\alpha$ in the three observations is also shown in the first panel. In these panels, velocities are calculated using rest energies of 5.89879, 6.40546, 6.70042, and 6.97317\,keV (from top to bottom), and green tick marks indicate the main line components. The observation at maximum Be star orbital velocity (drawn in red) shows blueshifted iron lines relative to the observation at minimum Be star orbital velocity (drawn in blue). Typical 1-$\sigma$ error bars are shown for the latter observation.  \label{fig:linesx}}
\end{figure}

All three methods provide statistically comparable fits to the data. We report the results of the first method in Table \ref{tab:rv}, as it relies on a more readily available modelling tool. The values shown in the table correspond to the five-channel binning; the results of the other binning strategies are consistent within the errors. All reported radial velocities are positive, which can be explained by two effects. First, line formation at the surface of the white dwarf leads to a gravitational redshift of $\sim$60\,\kms. Second, the rest wavelengths of the fluorescence lines depend on the ionisation stage of iron, with larger values for more ionised cases. If neutral iron is taken as a reference, as in the first model, this leads to apparent redshifts of up to 140--200\,\kms\ for K$\alpha_{2,1}$ lines of Fe\,{\sc viii}. Systematic redshifts can therefore indicate fluorescence from non-neutral iron. However, the observed fluorescence feature likely includes contributions from multiple ionisation stages of iron, which complicates the determination of an absolute reference and, consequently, of absolute radial velocities. Furthermore, uncertainties in atomic parameters remain, with differing assumptions and results amongst authors. Differences of 30--70\,\kms\ exist between theoretical and measured K$\alpha_{1,2}$  wavelengths of Fe\,{\sc i}, increasing to 90--140\,\kms\ for Fe\,{\sc viii} \citep{pal03}. To avoid such interpretation problems, Figure \ref{fig:lines} focuses on velocity shifts relative to the first observation of our campaign. 

Two sets of emission lines are associated with highly ionised iron. The first set corresponds to the Fe\,{\sc xxv} He$\alpha$ complex, composed of four lines near 6.7\,keV, while the second consists of the Fe\,{\sc xxvi} Lyman $\alpha$ doublet near 7\,keV. We fitted these lines using three approaches: (1) the Xspec model \texttt{bvapec}, representing emission from a low-density, optically thin, thermal plasma, with the iron abundance, plasma temperature, redshift (i.e. velocity), and broadening left as free parameters; (2) a set of four Gaussians with identical widths and velocities to fit only the Fe\,{\sc xxv} complex, with their strengths free to vary except for the two central intercombination lines (the strength of the line at 6.68232\,keV was set at 90\% of that at 6.66757\,keV); (3) a pair of Gaussians with identical widths and velocities, but independent strengths, to fit only the Fe\,{\sc xxvi} doublet. For the Gaussian models, we added a power law with zero slope to account for the continuum level. All three models provide comparably good fits to the data. Table \ref{tab:rv} reports the results of the first model, using the five-channel binning scheme. This model simultaneously fits both ionised line complexes, and thus yields more precise parameter estimates. Fig. \ref{fig:lines} shows the derived relative velocities (with green symbols) and line widths. 

\begin{figure}
  \begin{center}
    \includegraphics[width=8cm]{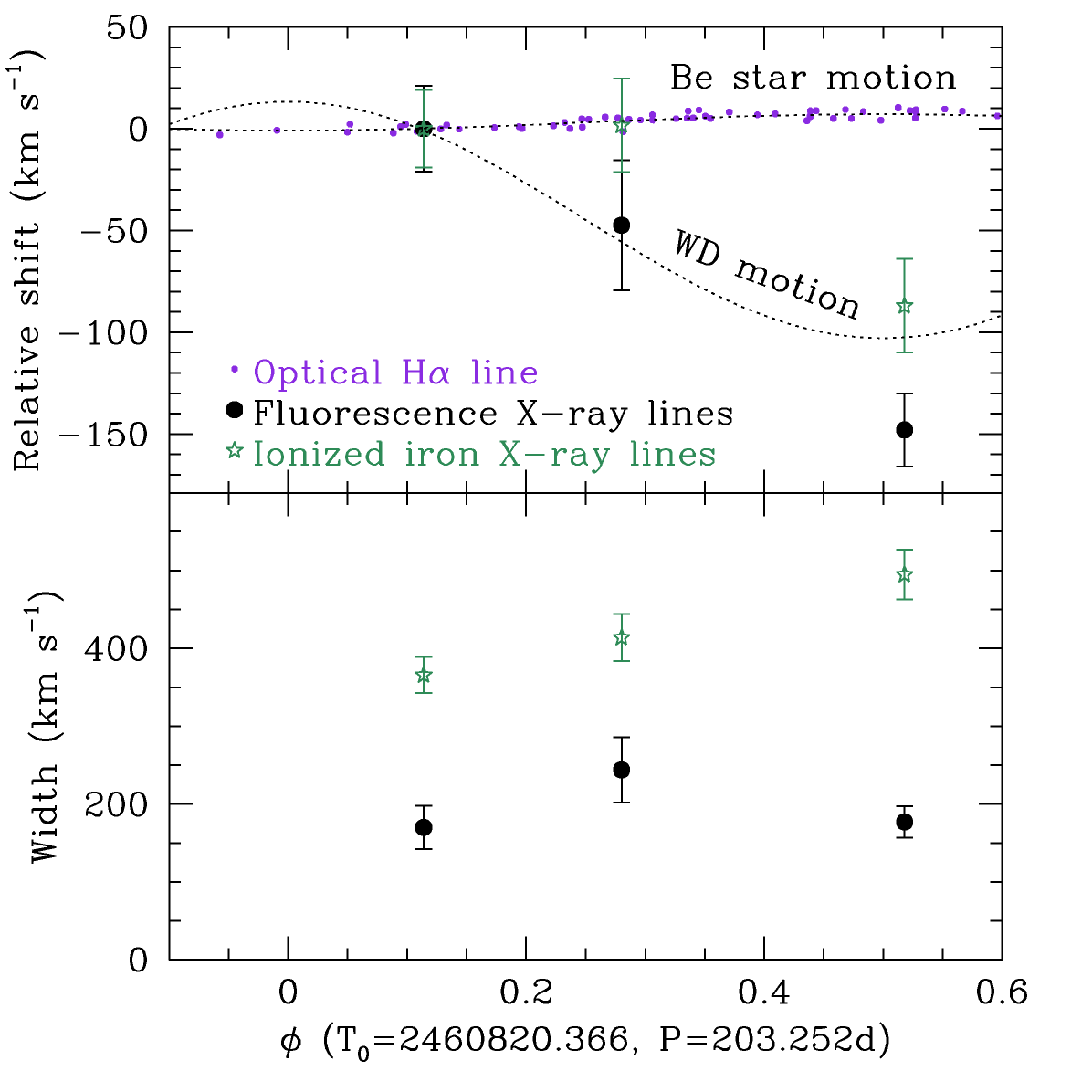}
  \end{center}  
  \caption{Shifts (top) and widths (bottom) of the iron features measure on \gc\ XRISM spectra. Fluorescence lines are shown as black dots, ionised lines with green stars. As in Table \ref{tab:rv}, the error bars are 1-$\sigma$ uncertainties. In the top panel, all velocities are relative to that at the first observing phase and are compared to the expected orbital motion of both the Be star and its white dwarf companion, demonstrating that the iron lines follow the white dwarf motion. For comparison, small violet points indicate velocities measured on the H$\alpha$ line, which is typically associated with the Be disc, at the time of the X-ray campaign \citep{naz25}. \label{fig:lines}}
\end{figure}

\begin{figure}
  \begin{center}
    \includegraphics[width=9cm]{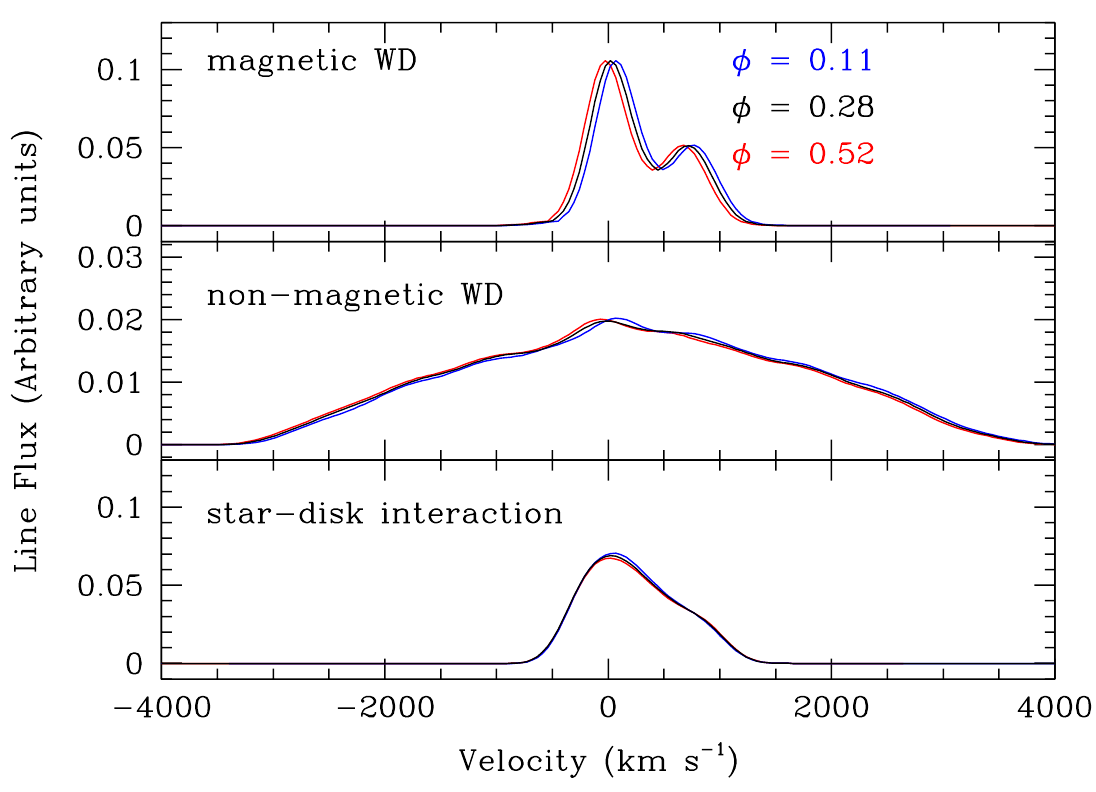}
  \end{center}  
  \caption{Line profile modelled at the {\it Resolve} resolution including all subcomponents of Fe\,{\sc iv} \citep{pal03}. Profiles are shown for the three orbital phases of the observations and three different scenarios: from top to bottom: accretion onto a magnetic white dwarf, accretion onto a non-magnetic white dwarf, and interaction between the Be star and its disc (see \citealt{rau25}). The colour convention follows Figs. \ref{fig:optlines} and \ref{fig:linesx}.  \label{fig:lines2}}
\end{figure}

Finally,  \citet{rau25} predicted the fluorescent line emission for the three plausible scenarios (Be star-disc interaction, non-magnetic accreting white dwarf, and magnetic accreting white dwarf). The primary X-ray source is described by an optically thin thermal plasma model of temperature 12.5\,keV. It illuminates several reservoirs of cool material: the Be decretion disc, the Be photosphere, the circumstellar environment of the white dwarf, and the white dwarf atmosphere. We discretised each medium into small cells and built the line profiles of individual fluorescent components by adding the contribution of each cell, accounting for its Doppler shift resulting from the cell velocity along the observer's sight-line. We modelled the cross-sections for the K-shell photo-ionisation, which precedes the fluorescent emission, using the formalism of \citet{Ver95}. We adopted the global structures of the Be circumstellar environment and the accretion structure around the white dwarf from smooth particle hydrodynamics models \citep{Ras25}. We described the density structures of the Be and white dwarf atmospheres using plane-parallel model atmospheres outside local thermodynamic equilibrium \citep{Hub21}. We treated the radiative transfer of fluorescent X-rays in these atmospheres with the Feautrier formalism \citep{Gar10}. We then obtained the synthetic profiles of the full Fe K$\alpha$ complex by convolving the individual line profiles with the energy distribution of each transition \citep{pal03}, accounting for their intrinsic strength. We also convolved with a Gaussian of 5\,eV full width at half maximum to represent the resolution of the {\it Resolve} instrument (Fig. \ref{fig:lines2}).

In the Be star-disc interaction scenario, the fluorescent line complex forms mostly in the innermost parts of the Be disc and follows the orbital motion of the Be star. Given the low amplitude of this orbital motion compared to the spectral resolution of the {\it Resolve} instrument, the observed line profile is essentially stationary. This contrasts with the situation in which the X-rays originate near the white dwarf. In these configurations, the line follows the white dwarf's orbital motion. In the non-magnetic white dwarf scenario, the fluorescence occurs mostly in the material in the inner parts of the white dwarf accretion disc. Because of the high rotational velocities of this matter, the ensuing fluorescent lines are expected to be very broad, with total widths of roughly 6000\,km\,s$^{-1}$ if the white dwarf has a mass near 1\,M$_{\odot}$ and a radius of 0.01\,R$_{\odot}$ (values well adapted to the case of \gc). For a magnetic white dwarf scenario, the magnetic field truncates the accretion disc, and the strongest contribution to the fluorescent line comes from reprocessing X-rays in the white dwarf's atmosphere (Fig. \ref{fig:mod}). The resulting synthetic line profile is much narrower than that when the accretion disc extends to the white dwarf surface. Furthermore, the fluorescence line is stronger when primary X-rays are emitted closer to the white dwarf surface \citep{rau25}.

\section{Results}
Figure \ref{fig:linesx} shows close-ups of the observed spectra, focusing on the three principal components of the iron complex at 6.4--7.0\,keV. The lines shift between observations. We measured these shifts using three different spectral binnings and three independent methods (see the previous section). All measurements appear mutually consistent, and Figure \ref{fig:lines} shows the evolution of line shifts and widths as a function of the orbital phase.

The radial velocity of the fluorescence complex decreases as the orbital phase changes from $\phi$=0.11 to 0.52. The velocity shift between these two extrema is $\Delta v=-148\pm28$\,\kms\ in the first method, representing a statistically significant detection at the 5.5 $\sigma$ level. The significance is slightly lower for the two-component models, due to fewer contributing lines, but remains well above 4 $\sigma$. The velocity shift measured on the ionised lines between quadratures is $\Delta v=-87\pm30$\,\kms, corresponding to a 3$\sigma$ detection. The trend is consistent in direction and amplitude with the velocity shift observed for the fluorescence lines. Additionally, we independently computed line centroids for each line compound (fluorescence, Fe\,{\sc xxv}, and Fe\,{\sc xxvi}) using the flux-weighted average velocity $\sum F_i v_i/\sum F_i$, where $F_i$ and $v_i$ are the flux and velocity of the $i$th pixel, respectively. The resulting centroids confirm the same trend.

In both cases, the velocity shifts exceed the systematic uncertainties from energy calibration and significantly deviate from constancy. We now compare them to orbital motions. Based on the most recent orbital solution derived from optical data \citep{naz25}, the Be star's radial velocity varies by only $\Delta v=+7$\,\kms\ over the same phase range. Although the compact companion does not appear in the optical spectra, we infer its expected motion from the Be star's mass function, the estimated mass of the Be star (13--16\,M$_{\odot}$), and the system inclination (42$^\circ$, \citealt{ste12}). This yields an expected velocity amplitude for the companion, over the same phase range, of $\Delta v \sim -100$\,\kms. The velocity shift observed for ionised lines is slightly smaller than this amplitude, whereas the fluorescence shift is slightly larger. Such deviations are statistically expected due to measurement uncertainties. The observed velocity evolution of the X-ray lines thus tracks the motion of the companion, not that of the Be star, providing strong evidence that both the hot thermal plasma (traced by the ionised lines) and the cooler reflecting material (traced by fluorescence lines) originate in material associated with the compact object.

The fluorescence feature requires significant line broadening to achieve satisfactory fits, with a best-fit Gaussian width of $\sigma^f_G \sim 200\pm30$\,\kms\ in the first method, differing from zero at more than the 5$\sigma$ level in all three observations. In the two-Gaussian models, the fitted widths are slightly larger, compensating for the lower number of line components. Furthermore, the three models assume symmetric Gaussian broadening; the success of the fits implies that there is no strong line asymmetry, consistent with theoretical expectations of fluorescence models \citep{rau25}. Across all models, the fitted line width appears marginally larger for the observation taken near conjunction ($\phi$=0.28, February 2025), but the difference corresponds to only a 1.5 $\sigma$ increase and is therefore not statistically significant within the current dataset. 

As suggested by previous grating observations for cooler plasma lines \citep{smi04,lop10,rau22}, significant line broadening is required to reproduce the ionised line profiles. The best-fit Gaussian width in the thermal plasma model is $\sigma^i_G \sim 425\pm30$\,\kms. This broadening is highly significant, exceeding the 10 $\sigma$  level in all three observations. As with the fluorescence, the models assume symmetric profiles, and the quality of the fits indicates that no strong line asymmetry is present. The observation at second quadrature ($\phi$=0.52, Feb 2025) yield a slightly larger line width, but the increase is marginal (between 1-3 $\sigma$, depending on the model) and cannot yet be considered significant. As for the fluorescence feature, higher-precision data will be required to confirm any phase-dependent variations in line broadening of these ionised lines.

Finally, we estimate the plasma temperature from the flux ratio of the Fe\,{\sc xxv} and Fe\,{\sc xxvi} lines. All three observations yield a consistent temperature of 100 MK (8.3\,keV, Table \ref{tab:rv}). This value is lower than the 125--150\,MK (11--13\,keV, \citealt{rau22}) typically derived from broadband spectral fits in earlier studies. However, the former is an ion temperature derived from line ratios, while the latter mostly reflects the electron temperature inferred from bremsstrahlung continuum. The discrepancy suggests a possible non-equilibrium plasma state in which ions and electrons are not thermally coupled, a scenario previously proposed for X-ray emission in shock-heated plasma in massive-star winds \citep{pol07}. Alternatively, Compton scattering or the presence of a multi-temperature plasma may also contribute to this difference.

\section{Discussion}
The XRISM results resolve the long-standing debate regarding the origin of \gc\ peculiar X-ray emission: the hard X-rays do not originate near the Be star, but are instead closely linked to its white dwarf companion. This result provides the first direct evidence in favour of the accreting white dwarf scenario \citep{mur86}. In this context, we note that the ionised features display equivalent widths of 92$\pm$4\,eV for Fe\,{\sc xxv} and 56$\pm$5\,eV for Fe\,{\sc xxvi}. Such values place \gc\ between intermediate polars and quiescent dwarf novae \citep{xu16}. This conclusion also holds for the strength of the fluorescence lines ($EW=43\pm2$\,eV) and the observed plasma temperature.

The accretion geometry onto the white dwarf depends critically on its magnetic field strength. For an actively accreting, non-magnetic white dwarf, both direct and fluorescent X-rays are emitted from the inner parts of the accretion disc, where high Keplerian velocities produce extremely broad emission lines. Models predict line widths of thousands of kilometres per second for a \gc-like configuration (see \citealt{rau25} and Fig. \ref{fig:lines2}) and such widths are observed, for example, in the dwarf nova SS\,Cygni \citep{ish25ss}. In contrast, our observations reveal only modest broadening, excluding such a configuration. In quiescence, the X-ray lines of dwarf novae may appear narrower ($\sim600$\,\kms\ for SS\,Cyg, \citealt{ish25ss}), but the geometry of the material near the white dwarf also differs, with direct accretion flows and fluorescence occurring mostly on the white dwarf surface. Our data cannot formally exclude such a configuration, but this scenario predicts outbursts, even if infrequent, and, no outburst has been detected so far for \gc\ or its analogues.

In magnetic white dwarfs, the magnetic field truncates the accretion disc and channels material along field lines onto the magnetic poles. In this configuration, shocks within the accretion column above the pole produce the primary X-rays, while fluorescence emission mostly originates from the white dwarf surface, with smaller contributions from the truncated disc, the Be disc, and material connecting the Be disc and the accretion structure (Figure \ref{fig:mod}). The observed line profiles agree with these expectations (\citealt{rau25}; compare Fig. \ref{fig:linesx} with Fig. \ref{fig:lines2}). They will be refined once more information on the white dwarf properties (rotation period and magnetic field strength) and the accretion rate become available. In addition, the strength of the fluorescence emission remains stable across all of our observations. To account for this strength through fluorescence from the white dwarf surface alone \citep{rau25}, most X-rays should be emitted at about 0.01\,R$_{\rm WD}$ above it. In the context of accretion onto a magnetic white dwarf, the widths of ionised lines likely arise from the velocity of shock-heated plasma that settles onto the surface of the white dwarf, but dedicated modelling is now needed for a thorough comparison. 

\begin{figure*}
  \begin{center}
    \includegraphics[width=12cm]{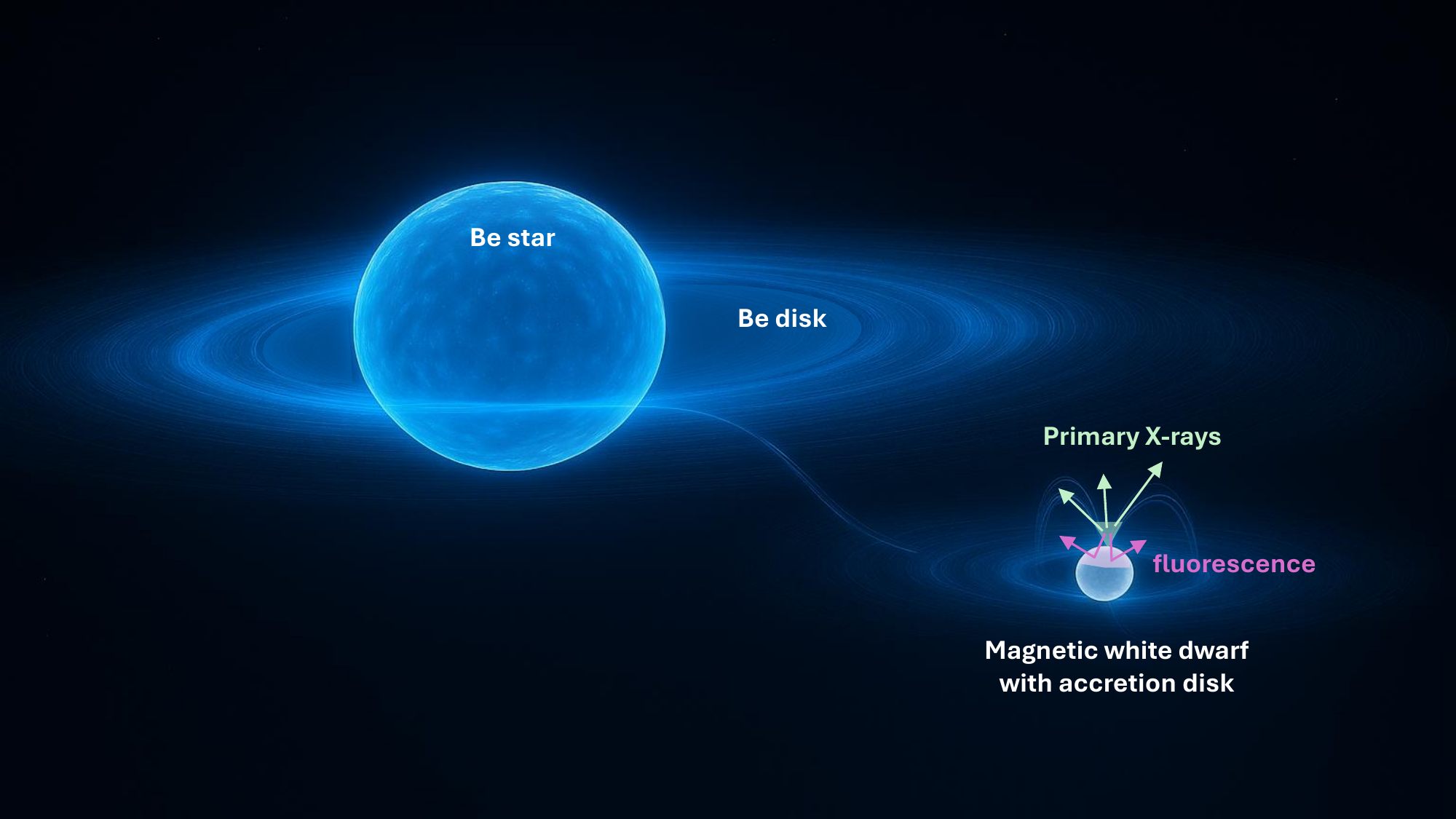}
  \end{center}  
  \caption{Artist view of \gc. The system consists of a massive Be star surrounded by a dense decretion disc, orbited by an accreting white dwarf. Material from the Be disk fuels the accretion disc of its companion. This accretion disc is likely truncated by the white dwarf's magnetic field, with accreted material then following magnetic field lines. Very hot plasma is generated in a shock above the magnetic pole, leading to X-ray emission. The X-rays then illuminate the white dwarf surface, the accreting material column, and the truncated accretion disk, which generates fluorescent X-rays. \label{fig:mod}}
\end{figure*}

How does the recognition of the origin of the hard X-ray emission of the \gc\ system fit within our knowledge of binary evolution? Population synthesis models, assuming various initial binary configurations and mass-transfer physics, predict that 50-70\% of Be binaries host white dwarf companions \citep{pol91,van97,rag01,sha14}, with orbital periods expected to be long (at least tens of days). These white dwarfs may represent either the direct products of the mass-transfer phase or the evolved remnants of stripped helium stars \citep{zhu23}. However, observational surveys have long lacked evidence for such systems -- a challenge that now appears resolved by the identification of an accreting white dwarf companion in \gc. Indeed, the distinctive X-ray properties shared by all \gc\ analogues make it very likely that they are all Be-white dwarf binaries. 

Nevertheless, this discrepancy may not be fully reconciled. Models \citep{sha14,zhu23,wan24} predict that Be stars paired with white dwarfs should predominantly have relatively low masses ($<10$\,M$_{\odot}$). Sensitive X-ray surveys \citep{naz20,naz23} reveal, however, that \gc\ analogues are confined to early spectral types (earlier than B3), with approximately half earlier than B1.5, corresponding to stellar masses $>10$\,M$_{\odot}$. Furthermore, the incidence of the \gc\ phenomenon reaches only $\sim$10\% amongst early-type Be stars in a distance-limited sample \citep{naz23}. These trends, which contrast with model expectations, point to a potential mismatch in current theoretical treatments of binary interactions, most likely in the parameters governing mass transfer.

This interpretation aligns with recently proposed revisions to mass-transfer efficiency based on observed stripped-star populations. \citet{lec25} showed that the observed properties of these systems disagree with predictions from the usual models, which use highly non-conservative mass transfer limited by the centrifugal barrier or the thermal timescale of the accretor. Instead, a high efficiency (60--80\%) of mass transfer should be used, a conclusion also reached by \citet{bao25}. Similarly, \citet{lab25b} report higher-than-predicted masses for the accretors. In parallel, \citet{pic25} studied a Be paired with a bloated stripped star and could not reproduce the system's observed properties (temperatures, luminosities, mass ratio, and period), despite investigating both case A and B mass transfer and implementing model adjustments (e.g. overshooting). Finally, \citet{xu25} examined the entire massive star population of the Small Magellanic Cloud and detected several discrepancies between observations and model predictions (e.g. number of Be X-ray binaries, Wolf-Rayet stars paired with OB stars, or OB stars paired with black holes), suggesting a lower merging rate and/or higher mass transfer efficiency. Together, these studies, including ours, strongly suggest that current binary models require revision. These revisions will certainly have a noticeable impact on our understanding of the full evolution of massive binaries, including their final supernovae and their compact endpoints.

Further theoretical investigation is also needed to understand the observed correlations (e.g. \citealt{mot15,smi16,rau22}) between X-ray variations and broadband optical and UV flux changes (typically attributed to variations in the inner Be disc) as well as the persistence of X-ray emission during epochs when the Be disc, which supplies material to the accretion structure, has largely dissipated \citep{naz22b}. Addressing this requires detailed hydrodynamical simulations, including radiative transfer, of Be-white dwarf binaries in \gc-like configurations. This task lies beyond the scope of this work. However, without models specifically dedicated to these Be-white dwarf systems, such features will remain difficult to interpret. Thus, revealing the true nature of \gc\ opens new avenues for research.

\begin{acknowledgements}
  The authors thank M. Loewenstein and F. S. Porter at GSFC for their advice in the energy gain calibration, M. Ishida at ISAS for sharing the results of SS Cygni prior to publication, and J. M. Miller at the University of Michigan for providing a part of the {\it Resolve} calibration data in the observation next to the first \gc\ observation. YN and GR acknowledge support from the FNRS and the university of Liège. This research made use of the JAXA's high-performance computing system JSS3. Support for SJG was provided by NASA through the SAO contract SV3-73016 to MIT for support of the CXC and Chandra Science Instruments (CXC is operated by SAO for and on behalf of NASA under contract NAS8-03060). SJG also received support from NASA through XRISM award number 80NSSC25K7499 issued by NASA. Finally, this research has made use of the CDS and NASA ADS services.
\end{acknowledgements}

{\bf Data availability. } The observational data will be publicly available in the JAXA and NASA archives (https://darts.isas.jaxa.jp/  and https://heasarc.gsfc.nasa.gov/ ) after the proprietary period ends.

\bibliographystyle{aa}
\bibliography{gCas_xrism.bib}

\end{document}